\documentclass{article}

\usepackage{graphicx}
\usepackage{subfig}
\usepackage{dcolumn}
\usepackage{bm}
\usepackage{amssymb}
\usepackage{amsmath}
\usepackage{amsfonts}
\usepackage{multirow}
\usepackage[usenames, dvipsnames]{color}
\usepackage{verbatim}
\usepackage{enumerate}
\usepackage{slashed}
\usepackage{xspace}
\usepackage[figuresright]{rotating}
\usepackage{listings}
\usepackage[baseurl=./]{hyperref}

\newcommand{\prog}[1]{{\sc #1}\xspace}
\newcommand{\code}[1]{{\tt #1}\xspace}
\newcommand{\codes}[1]{{\tt #1}s\xspace}
\newcommand{\SJ}[0]{\prog{SpartyJet}}
\newcommand{\FJ}[0]{\prog{FastJet}}

\title{\SJ 4.0 User's Manual}
\author{Pierre-Antoine Delsart, Kurtis L. Geerlings, Joey Huston,\\Brian T. Martin, and Christopher K. Vermilion}
\begin{document}
\maketitle
\tableofcontents

\section{Overview}

\SJ is a set of software tools for jet finding and analysis, built around the \FJ library of jet algorithms \cite{FastJet,FastJetWebsite}.  \SJ provides four key extensions to \FJ:  a simple Python interface to most \FJ features,  a powerful framework for building up modular analyses, extensive input file handling capabilities, and a graphical browser for viewing analysis output and creating new on-the-fly analyses.  Many of these capabilities rely on a \prog{ROOT}-based backend \cite{ROOT}.  Beyond finding jets, many jet tools in \SJ perform measurement of jet or event variables, referred to here as jet or event ``moments''.  These moments are available to subsequent tools and stored in the final output.  \SJ can be downloaded from HepForge at \url{http://projects.hepforge.org/spartyjet}.

Sec.~\ref{sec:example} walks through a simple example.  Sec.~\ref{f0:Installation} gives instruction on installation.  Sec.~\ref{f0:jetbuilder} describes the overall structure of a \SJ run.  Sec.~\ref{f0:JetTools} describes and enumerates the various jet tools available in \SJ.  Secs.~\ref{f0:Input} and \ref{f0:Output} describe \SJ's input and output facilities.  Finally, Sec.~\ref{f0:GUI} describes the capabilities of \SJ's graphical interface.


\section{A simple example}
\label{sec:example}

The simplest way to get a feel for how \SJ works is to consider an example; in this section we walk through the script \verb+examples_py/simpleExample.py+.  This script runs the anti-$k_\text{T}$ algorithm on the first 10 events listed in \verb+data/J1_Clusters.dat+, makes a simple measurement on the found jets, and stores the results.

\begin{itemize}
\item Load the libraries that are needed for the algorithms that you are running (you need to \code{source setup.sh} in the main directory first to set up environment variables):
\begin{lstlisting}[language=Python]
from  spartyjet import *
\end{lstlisting}

\item Create a \code{JetBuilder} object to manage \SJ analyses. The argument sets the output message level --- options are \{\code{DEBUG, INFO, WARNING, ERROR}\!\!\}.  Log messages throughout the code are tagged with a message level; messages with level lower than the current output level are suppressed.  Note that the \code{SpartyJet} namespace is aliased as \code{SJ}.
\begin{lstlisting}[language=Python]
builder = SJ.JetBuilder(SJ.INFO)
\end{lstlisting}

\item Create an input object with the name of the file containing the events, using the \code{getInputMaker} helper function.  This function guesses the type of input file based on the file extension.  Many types of input are available; see Section \ref{f0:Input}.
\begin{lstlisting}[language=Python]
input = getInputMaker('../data/J1_Clusters.dat')
builder.configure_input(input)
\end{lstlisting}

\item  Define the jet analyses that you want to run.  Most analyses begin with a jet algorithm, implemented in \FJ via the \code{FastJetFinder} tool.  See Section for \ref{FastJetFinder} for more options.  Note that the \code{fastjet} namespace is aliased as \code{fj}.
\begin{lstlisting}[language=Python]
name = 'AntiKt4'
alg = fj.antikt_algorithm
R = 0.4
antikt4Finder = SJ.FastJet.FastJetFinder(name, alg, R)
analysis = SJ.JetAnalysis(antikt4Finder)
\end{lstlisting}
 
\item  Pass analysis to your \code{JetBuilder}.   A sequence of jet tools defines a jet analysis.   A jet finder is one example of a \code{JetTool}; see Section \ref{f0:JetTools} for many more.
\begin{lstlisting}[language=Python]
builder.add_analysis(analysis)
\end{lstlisting}
 
 \item Insert another tool in the chain, in this case making a measurement of jets' angular moments in $\eta$ and $\phi$.  These will also be stored in the output \prog{ROOT} file.  \code{add\_jetTool()} can take a second argument, the name of an algorithm to add the tool to; otherwise the tool is added to all algorithms.  Tools can also be added directly to analyses via \code{JetAnalysis::add\_tool()}.
\begin{lstlisting}[language=Python]
builder.add_jetTool(SJ.EtaPhiMomentTool())
\end{lstlisting} 
 
\item  Configure (optional) simple text output for quick visual check of results.
\begin{lstlisting}[language=Python]
builder.add_text_output('../data/output/text_simple.dat')
\end{lstlisting}

\item Configure the Ntuple output by specifying the name of the tree and the \code{ROOT} file you want the data to be stored in.  This output can be manipulated via your own \code{ROOT} scripts or be viewed with the \SJ GUI. 
\begin{lstlisting}[language=Python]
builder.configure_output('SpartyJet_Tree','../data/output/simple.root')
\end{lstlisting}

\item Give the command to run the algorithms on the first 10
events 
\begin{lstlisting}[language=Python]
builder.process_events(10)
\end{lstlisting}

\end{itemize}

Running the script will process the first 10 events on the file specified in the input object and produce the \verb+.root+ file specified in \code{configure\_output}.  To view the output with the GUI, run
\begin{verbatim}
sparty data/output/simple.root
\end{verbatim}
(\verb+spartyjet/bin+ must be in your \verb+$PATH+.)  This is meant as a basic introduction, and there are many more functions than listed here.  See the other examples for more.

\section{Installation}
\label{f0:Installation}

\subsection{Requirements}

\SJ should build and run on any Unix-like operating system, including Mac OS X. A Cygwin build is presumably possible but not tested.  The standard build system is via Makefiles; there is also a CMake build system which requires CMake.  \SJ uses \prog{ROOT} extensively for input, output, and Python wrapping via PyROOT.  \SJ 4.0 has been tested with versions 5.30 and 5.32, but older versions may be OK.  To use the Python interface to \SJ you must have built \prog{ROOT} with PyROOT enabled.  We have tested with Python versions 2.6 and 2.7.  \SJ now makes heavy use of recent \FJ features, so \FJ version 3.0+ is required.  If you do not already have \FJ installed, \SJ can build an internal version.  If you want to use the StdHEP input facility, you will need a Fortran compiler; we've tested with gfortran.

\subsection{Compilation}

To compile:\\

\code{
\noindent
\# Set up ROOT such that root-config is in your path\\
\# for example source root/bin/thisroot.sh\\
cd spartyjet\\ 
source setup.sh\\
make\\
}

If you do not already have \FJ installed, do \verb+make fastjet+ and \verb+source setup.sh+ (again) before \verb+make+.  This builds  \FJ inside \SJ and sets the relevant environment variables.

\SJ has several building options to note:
\begin{itemize}

\item {\bf FastJet:} \SJ depends on \FJ for jet finding and some internal features, and the latest version is included in the \SJ distribution.  If you prefer to use your own installation, simply add \code{your-fastjet/bin} to the environmental variable \code{\$PATH} such that \code{fastjet-config} can be found.\\  
{\bf NOTE:} If you have linking problems between your version of \FJ and \SJ, either recompile your \FJ with the \code{--with-pic} option enabled before compilation, or have \SJ compile its own version.  Note also that to enable all \FJ plugins, you must pass the \code{--enable-allcxxplugins} flag to \code{configure}.  This is done by default for the built-in version.

\item {\bf StdHEP libraries:} These require a Fortran compiler and are automatically compiled if you have \code{gfortran}, \code{f77}, or \code{g77} in your  \code{\$PATH}. If you would like to try a different compiler, set the environmental variable \code{\$F77} to the compiler binary.

\item {\bf Pythia 6/8 interface:}  If you have \prog{ROOT} compiled with the Pythia 6 and/or Pythia 8 interfaces enabled, you can use this from within \SJ to generate events in Pythia and feed the output directly to \SJ.  To enable this, set the variable \code{\$PYTHIA6DIR} and/or \code{\$PYTHIA8DIR}.  If you do not have \code{PYTHIA} support in ROOT, you can add it by doing:\\
\code{cd \$ROOTSYS \\
./configure --enable-pythia6 --enable-pythia8\\ 
--with-pythia6-libdir=/my/pythia6/ \\
--with-pythia8-incdir=/my/pythia8145/include/ \\
--with-pythia8-libdir=/my/pythia8145/lib/\\
make}

\end{itemize}

Building \SJ creates a set of libraries in \verb+spartyjet/lib+ that you can load from a ROOT session or Python script, or you can link to to build an executable.

\begin{tabular}{l c l}
libs/libExternal.so&-&\FJ and other code \SJ depends on\\
libs/libJetCore.so&-&Core infrastructure\\
libs/libIO.so&-&Facilities for reading and writing a variety of file formats\\
libs/libFastJet.so&-&Tools that rely on \FJ, including jet finding\\
libs/libJetTools.so&-&Other JetTools\\
libs/libEventShape.so&-&Thrust and other event shapes\\
libs/libSpartyDisplay.so&-&\SJ GUI\\
libs/libExternalTools.so&-&A set of third-party jet tools, with wrappers\\
\end{tabular}

\subsection{Running}

Working examples of how to use \SJ can be found in the following directories:

\vspace{5pt}

\begin{tabular}{l c l}
\verb+spartyjet/examples_py+ &:& Python scripts (recommended) \\
\verb+spartyjet/examples_C+   &:& Compiled programs in C++ \\
\end {tabular}

\vspace{5pt}

The Python interface to \SJ is strongly preferred, and C++ access may be deprecated in a future release.  To use the Python scripts, some environment variables needs to be set, which can be accomplished via:
\begin{verbatim}
source setup.sh
\end{verbatim}
in the \verb+spartyjet/+ directory.  This exports the relevant paths to your \code{LD\_LIBRARY\_PATH} and sets the environment variable \code{SPARTYJETDIR}, which allows the \SJ Python modules and libraries to be accessible from any directory.  The relevant lines of \verb+setup.sh+ could also be copied into your shell's \verb+rc+ file, e.g. \verb+~/.bashrc+.  You will also need \verb+${ROOTSYS}/lib+ in your \verb+$PYTHONPATH+.  (This is necessary to use PyROOT.)

\subsection{CMake build}

\SJ now includes a  \href{http://cmake.org}{CMake} build system.  If you have CMake installed, you can build \SJ by creating a build directory (e.g., \verb+spartyjet/build+) and running:
\begin{verbatim}
cmake ..
make
\end{verbatim}
If you are using the built-in \FJ distribution, you need to run:
\begin{verbatim}
cmake ..
make fastjet
cmake ..
make
\end{verbatim}
To set environment variables correctly, you may need to run \verb+source setup.sh+ in the main directory before running \SJ programs.  In a future release, the CMake build system should allow natively building on Windows (not Cygwin), but this will likely require Windows builds for \SJ's dependencies.

CMake can generate project files for your IDE of choice, e.g., to build an Xcode project just do
\begin{verbatim}
cmake .. -G Xcode
\end{verbatim}


\section{\code{JetBuilder}}
\label{f0:jetbuilder}

\begin{figure}[htbp]
\begin{center}
\includegraphics[width=1.2\linewidth]{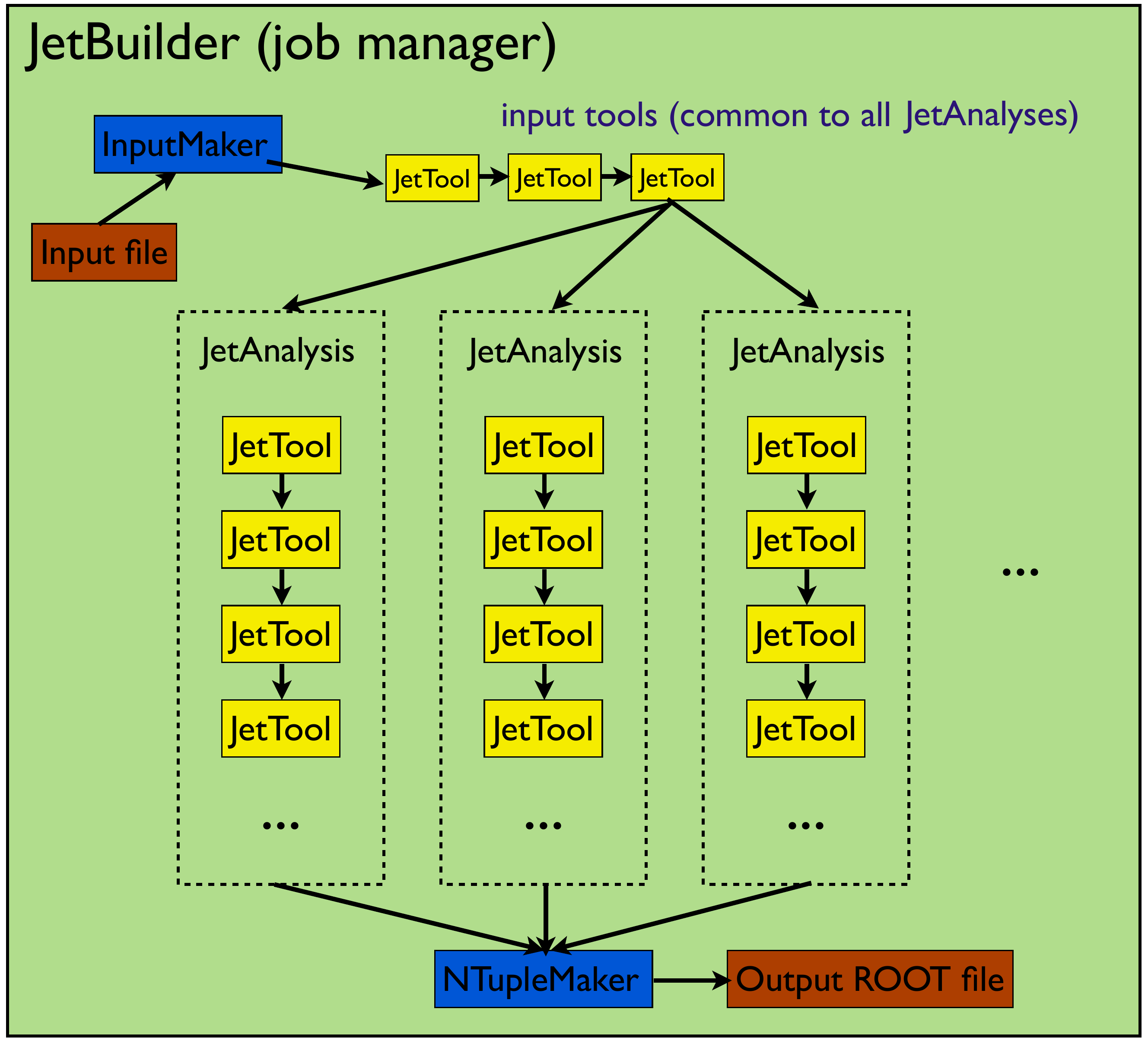}
\end{center}
\caption{The structure of a SpartyJet analysis.  A \code{JetBuilder} gets input from a file using an \code{InputMaker}, passes the input through a sequence of input \codes{JetTool}, and passes the output to several \code{JetAnalyses} (labeled JetAlgorithms in this diagram).  \code{JetAnalyses} consist of a sequence of \codes{JetTool} that are run sequentially.  The output of each \code{JetAnalysis} is passed to an \code{NTupleMaker} which stores the final jets and accompanying moments in a file. }
\label{fig:JetBuilder}
\end{figure}

\code{JetBuilder} is the job manager for \SJ.  \code{JetBuilder} takes the input from an \code{InputMaker} (see Section \ref{f0:Input}) and passes it through a set of \code{JetAnalyses}.  A \code{JetAnalysis} is made up of a sequence of \codes{JetTool}.  The final list of jets, and associated moments, is passed to an \code{NtupleMaker}, which prepares the output.  This is shown schematically in Fig.~\ref{fig:JetBuilder}.

\subsection{Input Functions}
{\bf Example:} \verb+examples_py/inputExample.py+

Input is passed to \code{JetBuilder} via:
\begin{lstlisting}[language=Python]
builder.configure_input(input, saveInput=True)
\end{lstlisting}
where input can be any class deriving from \code{InputMaker}.  See Section \ref{f0:Input} for examples.  The \verb+saveInput+ flag, true by default, determines whether to save the input particles in the output file.  Saving input particles is needed for most event displays but not for plotting run variables (jet mass, moments, etc.).

\subsection{\code{JetAnalysis} Options}

\code{JetAnalyses} are blocks of code that act on a set of \codes{Jet} --- implemented as a \code{JetCollection}.  A set of \codes{JetTool} forms a \code{JetAnalysis}, and a \code{JetBuilder} holds a set of \code{JetAnalyses} that are each run on its inputs.  Examples of specific tools can be found in Sec.~\ref{f0:JetTools}. 

\code{JetAnalyses} can be added to a run via:
\begin{lstlisting}[language=Python]
builder.add_analysis(analyis)
\end{lstlisting}

Alternatively, analyses can be set up implicitly by just passing the \code{JetBuilder} a \code{JetTool}:
\begin{lstlisting}[language=Python]
builder.add_default_analysis(tool)
\end{lstlisting}
where \code{tool} is typically a jet finder, which forms the basis of the tool chain.  Default tools are added before and after --- currently just negative energy correctors, if this option is enabled (it is not by default).  To enable negative energy correction, call
\begin{lstlisting}[language=Python]
builder.do_correct_neg_energy(True)
\end{lstlisting}
This flips the sign of the energy of any negative-energy input particles, then after jet finding adjusts jets' energies accordingly.

\codes{JetTool} can be added  directly to analyses, or through the \code{JetBuilder}:
\begin{lstlisting}[language=Python]
# add a tool a specific analysis
analysis.add_tool(tool)
# add a tool to all analyses registered with builder
builder.add_jetTool(tool)
# add a tool to a specific analysis via builder
builder.add_jetTool(tool, 'AnalysisName')
\end{lstlisting}
Tools can be added to the front of the sequence with:
\begin{lstlisting}[language=Python]
analysis.add_tool_front(tool)
# or
builder.add_jetTool_front(tool)
\end{lstlisting}

Analyses are passed to the \code{JetBuilder} as pointers, so tools can be added to them at any time before calling \code{JetBuilder::process\_events()}.

A single sequence of \codes{JetTool} is run on the input particles before they are passed to each \code{JetAnalysis}.  This sequence starts empty; to add to it use:
\begin{lstlisting}[language=Python]
builder.add_jetTool_input(tool)
\end{lstlisting}
This is useful for initial tools like $\eta$ cuts or detector simulation that are common to all algorithms.

\subsection{Output options}
\SJ by default retains information about each jet's constituents.  \code{JetBuilder::add\_analysis()} and \code{JetBuilder::add\_default\_analysis()} each take a second bool argument that determines whether to save constituent information for each jet:
\begin{lstlisting}[language=Python]
builder.add_analysis(analysis, withIndex=True)
builder.add_default_analysis(tool, withIndex=True)
\end{lstlisting}
The default is \verb+True+.  For this to be useful in the output, the input should be set up to store input particles in the output file as well.  The full recombination history is stored in memory, but storage to disk has not yet been implemented.  We hope to have this available in a future release.

A third option can be passed to \code{JetBuilder::add\_analysis()}, which if \verb+True+ tells the builder to only save jet and event moments, not jet variables like $\eta$, $p_T$, etc.  This can be useful to save space, or if only event-shape tools are used.  In fact, \code{JetBuilder} has a method \code{add\_eventshape\_analysis(analysis)} that is simply an alias to \code{add\_analysis(analysis, False, True)}.

A text file output option tells \code{JetBuilder} to produce an easily-readable text file that contains a list of all jets found from all algorithms for all events.  To turn on the text file output, you must call \code{builder.add\_text\_output()}, and pass it the filename you want to create. For example:
\begin{lstlisting}[language=Python]
builder.add_text_output('../data/output/text_output.dat')
\end{lstlisting}

The main output of \SJ is a \prog{ROOT} file holding input particles, jet momenta, and jet and event moments.  The format of this file as well as some additional \code{JetBuilder} options controlling it, are described in Sec.~\ref{f0:Output}.

\subsection{Minimum Bias Overlay}
{\bf Example:} \verb+examples_py/overlayExample.py+

\code{JetBuilder} allows the user to add minimum bias (MB) events to the signal events to study the effects of pileup.  To enable MB events, one must:
\begin{enumerate}
\item Create another input object:
\begin{lstlisting}[language=Python]
MBinput = SJ.StdTextInput('../data/MB_Clusters.dat')
\end{lstlisting}
\item Tell \code{JetBuilder} to add \code{n} MB events to each signal event:
\begin{lstlisting}[language=Python]
builder.add_minbias_events(n, MBinput, poisson=False)
\end{lstlisting}
\end{enumerate}
\code{JetBuilder} will start at the beginning of the MB data file and read the first \code{n} events for the first data event, then the second \code{n} events for the second data event, so on.  When the end of the MB file is found, it will simply continue from the beginning.   If the third, optional, argument is \verb+True+, the \code{JetBuilder} will draw the number of MB events from a Poisson distribution (default is \verb+False+).  In this case, \code{n} is the Poisson mean or expected number of MB events.

\subsection{Options controlling screen output}

\SJ can be configured to produce screen output at varying levels of verbosity.  The most important point of control is the global message level, which can be one of  \{\code{DEBUG, INFO, WARNING, ERROR}\!\!\}.  Each message produced in the course has an associated level, and will be printed to the screen if its level is at least as high as the global level: e.g., if the global level is \code{WARNING}, \code{WARNING} and \code{ERROR} messages will be printed.  The global message level can be set in the \code{JetBuilder} constructor, or later via the \code{JetBuilder::set\_message\_level()} method:
\begin{lstlisting}[language=Python]
builder = SJ.JetBuilder(SJ.INFO)
# or
builder.set_message_level(SJ.WARNING)
# equivalent to set_message_level(SJ.ERROR):
builder.silent_mode()
\end{lstlisting}

\SJ will report on the progress of the run by printing ``Processed Event: N'' lines, by default every event.  You can change the frequency via:
\begin{lstlisting}[language=Python]
# prints every 100 events
builder.print_event_every(100)
\end{lstlisting}

\subsection{Running}

Once all analyses have been set up and all options loaded, it's time to run.  To run over all events in the input, simply do
\begin{lstlisting}[language=Python]
builder.process_events()
\end{lstlisting}
You can also specify how many events to process, and what event number to begin at:
\begin{lstlisting}[language=Python]
# process 100 events, starting at the 250th
builder.process_events(100, 250)
# alias to process_events(1, N):
builder.process_one_event(N)
\end{lstlisting}

\section{\codes{JetTool}}
\label{f0:JetTools}

\codes{JetTool} are blocks of code that act on a \code{JetCollection} for every event.  They generally perform one or more of the following functions:
\begin{itemize}
\item Find jets
\item Add or remove jets from the list.
\item Modify the jets themselves.
\item Add information about each jet as a \code{JetMoment}.
\item Add information about each event as an \code{EventMoment}.
\end{itemize}
Jet tools can be added either before or after the primary jet finder (e.g., the $k_\text{T}$ algorithm) in the \code{JetAnalysis}.
This section describes the set of \codes{JetTool} shipped with \SJ.  The relevant definitions can be found in \verb+JetTools/+ (core \SJ tools), \verb+FastJetTools+ (wrappers around \FJ tools), and \verb+ExternalTools+ (\FJ-based tools provided by third parties).

\SJ and \FJ have evolved alongside each other, and many features formerly implemented natively in \SJ are now outsourced to \FJ algorithms or tools.  We have endeavored to keep pace, but realize there is still some duplication, especially with the suite of tools available in \FJ 3.0.  Where available, \FJ-based tools are the preferred means of performing jet finding (\code{fastjet::ClusterSequence}), selection (\code{fastjet::Selector}), manipulation (\code{fastjet::Transformer}), and measurement (\code{fastjet::FunctionOfPseudoJet}).  The improved wrapping of \FJ code in \SJ 4.0 should enable essentially any \FJ tool to be used within \SJ.  We expect that future development of jet tools will be done within the context of \FJ, with development in \SJ focused on the Python interface, I/O facilities, and graphical interaction.

As noted in \cite{FastJetManual}, the planned future development of \FJ includes a ``contrib'' project of externally written code.  We envision that many external tools currently provided with \SJ will migrate to this space when it becomes available.  In the meantime, see \verb+ExternalTools/README+ for details on the origins and authorship of the tools in that directory.  To add your own tools to \SJ, you will need to either create a shared library that a) links against compiled C++ code or b) is loaded by ROOT.  See the comments in \verb+UserPlugins/Makefile+ for more information on creating such a library, or including your own code in \SJ's build.  \verb+UserPlugins/ExamplePlugin+ gives an example.

Note: In this section we sometimes switch from Python to C++ syntax for clarity of argument types.

\subsection{Jet finders}
\label{FastJetFinder}
All jet finding is done via \FJ jet algorithms.  The basic wrapper is defined in \verb+FastJetTools/FastJetFinder.hh+.  There are two basic operations: jet finding, and reclustering.

\begin{description}
\item \code{FastJetFinder}:
Finds jets using \FJ algorithms.  It has two constructors:
\begin{lstlisting}
FastJetFinder(string name, fastjet::JetAlgorithm alg, double R=0.4, bool area=false)
FastJetFinder(fastjet::JetDefinition, string name, bool area=false)
\end{lstlisting}
The second form allows you to pass your own \code{fastjet::JetDefinition}, including a plugin algorithm; see \verb+FJExample.py+ for examples.  By default, a \code{FastJetFinder} finds inclusive jets with $p_T > 5$ GeV.  This can be modified via \code{set\_ptmin(double)}, \code{set\_dcut(double)}, and \code{set\_njets(int)}.  The last two correspond to \emph{exclusive} jet finding; note that \code{dcut} has dimension GeV$^2$.  Setting any of these overrides the others.

\item \code{FastJetRecluster}:  Similar to \code{FastJetFinder}, but uses a given jet algorithm to recluster the constituents of each jet in an event.  Has two constructors, with the same arguments as \code{FastJetFinder}.
\end{description}

The full set of default algorithms from \FJ is available.  To use the included but optional plugin algorithms, uncomment the relevant lines in \verb+External/ExternalLinkDef.hpp+ and recompile \SJ.

Two additional plugin algorithms are available in the \verb+ExternalTools+ directory:

\begin{description}
\item \code{NjettinessPlugin}: Finds jets by minimizing the $N$-jettiness measure \cite{Njettiness, NSub2} over the event with a fixed number ($N$) of axes.
\item \code{QjetsPlugin}: Implements the ``Qjets'' algorithm of \cite{Qjets}.
\end{description}

\subsection{Selectors}
These tools allow the user to remove some Jets from an input/output JetCollection.  Most selectors can now be formed with the standard \FJ\,\codes{Selector} via the \code{SelectorTool} class:
\begin{lstlisting}[language=Python]
selector = fj.SelectorPtMin(200.0) * fj.SelectorNHardest(2)
\end{lstlisting}
See the \FJ manual or \verb+fastjet-install/include/fastjet/Selector.hh+ for the full set of available selectors.  The compound assignment operators \code{\&=} and \code{|=} are supported, as well as (via some duck punching in the \SJ Python wrapper) the binary operations \code{*}, \code{\&\&}, and \code{||}.   Note that \code{(s1 * s2)} applies \code{s2}, then \code{s1}; \code{(s1 \&\& s2)} applies both separately.  The result is the same iff the selectors commute.

Many native \SJ selectors are defined in \verb+JetTools/JetSelectorTool.hh+; the ones that do not duplicate a \FJ selector (and are thus not deprecated) are:

\begin{description}
\item \code{JetInputPdgIdSelectorTool(std::vector<int> pdgIds)}: Removes input ``jets'' with given PDG IDs.  (Useful when the input jets are just single particles from a Monte Carlo where leptons, neutrinos, etc. are included.)  The input file must have included PDG IDs.
\item \code{JetMomentSelectorTool<T>(std::string momentName, T min, T max)}: Finds the given jet moment (calculated by another tool) and requires it be within (min, max).  T can be any type supporting less-than comparison.
\end{description}

\subsection{Transformers}
\FJ 3.0 provides a common base class for jet manipulation: \code{fastjet::Transformer}.  Transformers can remove particles (e.g., filtering), re-arrange substructure, or tag/reject jets.  To add a given transformer to an analysis, use the \code{TransformerTool}:

\begin{lstlisting}[language=Python]
Trimmed = SJ.JetAnalysis([... some jet finder ...])
trimmer = fj.Filter(0.35, fj.SelectorPtFractionMin(0.03))
Trimmed.add_tool(SJ.FastJet.TransformerTool("TrimmerTool", trimmer))
\end{lstlisting}

In C++ code any class deriving from \code{fastjet::Transformer} is acceptable; in Python a \prog{ROOT} dictionary must be generated.  See the comments in \verb+UserPlugins/Makefile+ for more information on including your own classes in \SJ.  The currently available set of \codes{Transformer} is:

\FJ native:
\begin{description}
\item \code{Boost}:  Boost a jet to the rest frame of a reference jet.
\item \code{CASubJetTagger}:  Versatile and generic substructure identification for Cambridge/Aachen jets.
\item \code{Filter}:  Recluster a jet with a smaller $R$ and only keep subjets passing a given criterion \cite{BDRS}.
\item \code{GridMedianBackgroundEstimator}: Two different ways of estimating background radiation.
\item \code{JetMedianBackgroundEstimator}
\item \code{MassDropTagger}: Tagger that peels off soft radiation until a splitting is found where the two parents have significantly less mass than the child \cite{BDRS}.
\item \code{RestFrameNSubjettinessTagger}: Implements the rest-frame version of the $N$-subjettiness jet shape \cite{RestFrameNSub}.
\item \code{Subtractor}: Subtracts estimated background energy.
\item \code{JHTopTagger}: The Johns Hopkins top tagger \cite{JH}.
\item \code{Unboost}: Unboost a jet from the rest frame of a reference jet.
\item \code{Pruner}: Rebuilds a jet with a new cluster sequence, vetoing soft, large-angle mergings \cite{Prune1, Prune2}.
\end{description}

External tools included with \SJ:
\begin{description}
\item \code{HEPTopTagger}: An implementation of the \code{HEPTopTagger} of \cite{HEP1, HEP2}, provided by the authors.
\item \code{CMSTopTagger}: An implementation of the ``CMS'' top tagger \cite{CMS1, CMS2}, written by one of the \SJ authors, based on \code{JHTopTagger}.
\item \code{TopTaggerDipolarityTool}: Adds dipolarity \cite{Dipolarity} to any top tagger, running the tagger, then storing the dipolarity measured on its substructure.
\end{description}

The \FJ-native transformers are described further in the \FJ documentation.  Examples of using many of these transformers can be found in \verb+examples_py/FJToolExample.py+ and \verb+examples_py/Boost2011.py+.

\subsection{Jet and event moment tools}

These tools calculate moments for each jet or event, which can be any type that ROOT knows how to store, most commonly \codes{double} or \codes{int}.  The generic \code{JetMomentTool} and \code{EventMomentTool} calculate and store any user-implemented \code{JetMoment<T>} or \code{EventMoment<T>} object.  See \verb+JetTools/JetMomentTool.hh+ for some specific examples, and \verb+FJToolExample.py+ for moment tools in action.

The \FJ 3 base class \code{FunctionOfPseudoJet<T>} provides a common interface for jet measurements.  This is wrapped in \SJ by the \code{PseudoJetMomentTool<T>} class.  The template argument \code{T} gives the type of the function output.

\begin{lstlisting}[language=Python]
# An example of using a FunctionOfPseudoJet measurement
nsub3 = fj.Nsubjettiness(3, Njettiness.onepass_kt_axes, 1.0, 1.0)
AntiKt10.add_tool(SJ.FastJet.PseudoJetMomentTool(double)(nsub3, "tau3"))
\end{lstlisting}

Most jet and event moments in \SJ are of type \code{Double32\_t}.  This is a \prog{ROOT}-defined type that behaves like a \code{double} in memory but is stored as \code{float}; we feel this is a reasonable compromise between precision mid-calculation and saving storage space.  In the Python package \code{ROOT.Double32\_t} is imported as simply \code{double}, as in the above example.

The following \codes{FunctionOfPseudoJet} are available, all defined in \verb+spartyjet/ExternalTools+.  Examples of their use can be found in \verb+examples_py/Boost2011.py+.

\begin{description}
\item \code{MinMassFunction}: Finds three exclusive subjets, then measures the smallest invariant mass between any two, as in the CMS top tagger \cite{CMS1,CMS2}.
\item \code{zCellFunction}: Measures the $z_\text{cell}$ variable of the ``Thaler and Wang'' top tagger \cite{ThalerWang}.
\item \code{zCutFunction}: Measures the $z_\text{cut}^i$ of the ATLAS top tagger \cite{ATLAS}.
\item \code{Nsubjettiness}: Measures the version of $N$-subjettiness described in \cite{NSub}.
\end{description}

The following are other moment-storing tools available.  They may be phased out in favor of more explicitly \FJ-based versions.

\begin{description}
\item{\code{HullMomentTool}}:
This tool finds the convex hull enclosing each jet and saves the hull length and area as \code{Hull} and {HullA} respectively.

\item{\code{EtaPhiMomentTool}}:
This tool calculates angular second moments in $\eta$ and $\phi$ of each jet stores them as \code{M2eta} and \code{M2phi}.

\item{\code{PtDensityTool}}:
This tool calculates each event's $p_{T}$ density using \FJ.  It does this by finding all the jets in the event with no minimum $p_{T}$ requirement.  It then extracts the $p_T$ density from these jets by selecting the mean $p_T$ density for each bin in $\eta$.  These $p_T$ densities are stored as \code{ptDensity} and the $\eta$ bin limits are stored as \code{ptDensityBins}.

\item{\code{JetAreaCorrectionTool}}:
This tool uses the area of each jet and the pTDensity found by the \code{PtDensityTool} to calculate a correction (stored as jet moment \code{JetAreaCorr}) to the jet's $p_{T}$.

\item{\code{YSplitterTool}}:
\label{sec:ysplitter}
This tool uses \FJ to calculate the $y$ values associated with a set of recombinations.  In this implementation, \SJ will run a \FJ algorithm of the user's choice on the constituents of a given jet. It can be called with either of the following constructors:
\begin{lstlisting}
YSplitterTool(float R, fastjet::JetAlgorithm alg, int ny, int njet)
YSplitterTool(fastjet::JetDefinition *jet_def, int ny, int njet)
\end{lstlisting}
where \code{njet} is the number of jets for which the $y$ values will be calculated and \code{ny} is the number of $y$ values to calculate for each jet.
\end{description}

\subsection{Miscellaneous tools}

\begin{description}
\item{\code{ForkToolParent} and \code{ForkToolChild}}:
This pair of tools allows the forking of \code{JetTool} chains.  \code{ForkToolParent} merely saves a copy of its input.  A \code{ForkToolChild} is associated with a specific parent, and it reads in the \code{JetCollection} saved by its parent.  This allows, for example, one jet algorithm to be run combined with several different jet-modifying tool chains for comparison.  See \verb+FJToolExample.py+ for a usage example.

\item{\code{CalorimeterSimTool}}:
This tool applies a very simple calorimeter simulation to its input jets.  Inputs are sorted into calorimeter cells on a specified $\eta$--$\phi$ grid.  For each non-empty cell, a massless output particle is created with the direction of the cell and the total energy of all particles in the cell.

\item{\code{RadialSmearingTool}}:
This tool wraps Peter Loch's \code{DetectorModel} code for calorimeter simulation.  For more information, see Peter's \href{http://atlas.physics.arizona.edu/~loch/index.html}{website}.  A usage example is given in \verb+Boost2011.py+.

\item{\code{JetNegEnergyTool}}:
This tool is meant to be run twice: once before jet finding and once afterward.  On first run, the tool finds and stores all input particles with negative energy.  For each such particle it inverts the energy to be positive.  On second run, the \code{JetNegEnergyTool} loops over jets, and for each constituent that initially had a negative energy, it corrects the jet energy by subtracting twice the constituent's (positive) energy.  The \code{JetBuilder} method \code{do\_correct\_neg\_energy(true)} inserts this pair of tools before and after all jet finders added with \code{add\_default\_analysis(tool)}; by default this is not done.

\item{\code{EConversionTool}}:
This tool simply converts the units of all the jets between MeV and GeV.  The user can convert to arbitrary units as well.

\item \code{HardProcessMatchTool}:
Assuming the input file includes information about the hard scattering (this is true of HepMC files, e.g.), finds the closest hard parton for each jet and stores the $\Delta R$ distance as a jet moment.

\item \code{AngularCorrelationTool}:
Measures the angular correlation functions described in \cite{ACF}, storing $R_n$ and $m_n$ for each peak found up to three as jet moments.

\item \code{QjetsTool}:
Runs the ``Qjets'' plugin algorithm repeatedly on each jet, storing the average pruned jet mass and its ``volatility'', as described in \cite{Qjets}.

\item{\code{WTaggerTool}}:
This tool wraps the $W$-tagging method of \cite{Wtag}, which includes a large number of substructure and mass cuts.  The cuts are taken from data files in \verb+external/wtag-1.00/data+.  So far there is no way to re-train the cuts from within \SJ --- the idea is to take a pre-trained $W$-tagging method and plug it into a \SJ analysis.  Since the tagger is taken ``out-of-the-box'', there are no input parameters:

\begin{lstlisting}
WTaggerTool()
\end{lstlisting}

\end{description}


\section{Input}
\label{f0:Input}

All \SJ jobs need an \code{InputMaker} object to read input from some data file and prepare a list of four-vectors for \code{JetAnalyses} to process.  There are several types of \codes{InputMaker} available.  An example of the implementation for each type of input can be seen in \verb+examples_py/inputExample.py+.  For Python scripts, the top-level module \verb+python/spartyjet/__init__.py+ defines the helper function \code{getInputMaker(fileName)} which will create the appropriate \code{InputMaker} by looking at the filename extension.

\subsection{NtupleInputMaker}
{\bf Sample:} \verb+data/J2_clusters.root+

This form of input reads \prog{ROOT} files.  This \code{InputMaker} requires the components of the input 4-vectors to be stored in separate branches of a \prog{ROOT} \code{TTree}.  The following definitions are supported:
\begin{itemize}
\item \code{px, py, pz, E}
\item \code{(psuedo)rapidity, phi, pt, E}
\item \code{(psuedo)rapidity, phi, pt, m}
\end{itemize}
You need only specify how the information is stored (array or vector / float or double) and the names of the branches.

As an example, to set up an \code{NtupleInputMaker} to read the file \verb+data/J2_clusters.root+,
if you don't know how variables are internally stored in your ntuple, do the following:
\begin{verbatim} 
root -l data/J2_clusters.root
root [0] clusterTree->MakeClass("test");
\end{verbatim}
Open the file test.h and check to see how the variables are stored.  In this example, we see lines like:
\begin{lstlisting} 
vector<float> *Cluster_eta;
\end{lstlisting}
indicating that our 4-vectors are stored in vectors of floats.  Now to configure \SJ to accept this, we need to:

\begin{itemize}
\item Create an \code{NtupleInputMaker} of the correct type: (in our case: \code{vector<float>} for (\code{eta, phi, pt, E})).
\begin{lstlisting}[language=Python]
input = SJ.NtupleInputMaker(SJ.NtupleInputMaker.EtaPhiPtE_vector_float)
\end{lstlisting}
For full list of Input codes see: 
\verb+JetCore/InputMaker_Ntuple.hh+

\item Configure the names of the \code{TBranches}
\begin{lstlisting}[language=Python]
input.set_prefix('Cluster_')
input.set_n_name('N')
input.set_variables('eta','phi','p_T','e')
input.setFileTree('../data/J2_clusters.root', 'clusterTree')
\end{lstlisting}

\item Specify if input is massless, only useable in \code{eta,phi,pt,E} mode (if true, \code{pt} is ignored):
\begin{lstlisting}[language=Python]
input.set_masslessMode(True)
\end{lstlisting}

\item Set the input file and tree names:
\begin{lstlisting}[language=Python]
input.setFileTree('../data/J2_clusters.root', 'clusterTree')
\end{lstlisting}
\end{itemize}

{\bf Python Shortcut:}

To allow \SJ to configure your Ntuple using some assumptions use the following helper function:
\begin{lstlisting}[language=Python]
input = createNtupleInputMaker('../data/J2_clusters.root', inputprefix='Cluster')
\end{lstlisting}

\subsubsection*{DelphesInput}

\code{DelphesInputMaker} is a minor extension of \code{NTupleInputMaker} that reads the \prog{ROOT} files produced by the detector simulator \prog{Delphes}.  Only calorimeter cells are read in.

\subsection{StdTextInput}
{\bf Sample: \verb+data/J1_Clusters.dat+}

This form of input reads ASCII files.  To separate events, put one of the following lines between the events:
\begin{verbatim}
.Event
.event
N
n
\end{verbatim}
(only the .E or .e is important in the first two).

The form of the four vectors should be:
\begin{verbatim}
E px py pz
\end{verbatim}

This input is configured simply with:
\begin{lstlisting}[language=Python]
input = SJ.StdTextInput('../data/J1_Clusters.dat')
\end{lstlisting}

If the form is the opposite (px py pz E), then call the function
\begin{lstlisting}[language=Python]
input.invert_input_order(True)
\end{lstlisting}
and it will be read in properly.

An example of this input can be seen in 
\verb+data/J1_Clusters.dat+
 
\subsection{StdHepInput }
{\bf Sample:} \verb+data/ttbar_smallrun_pythia_events.hep+

This form of input reads StdHEP format XDR files. It will look for
particles with the status code of 1 (final state).  It is also able extract the PDG ID code for each particle from the input data to allow further filtering and matching.  Access to intermediate particles such as the participants in the hard scattering, or $B$ hadrons from $b$ quark decays, should be possible in the near future.

This input is configured simply with:
\begin{lstlisting}[language=Python]
SJ.StdHepInput('../data/ttbar_smallrun_pythia_events.hep')
\end{lstlisting}

NOTE: To read StdHep files, you must enable StdHep compilation as explained in Section \ref{f0:Installation}.

\subsection{CalchepPartonTextInput}
{\bf Sample:} \verb+data/gg_ggg_events.dat+

This form of input reads output from CalcHEP. It reads in the number of
initial and final state particles, and then for each event saves only
the information for the final state particles.

This input is configured simply with:
\begin{lstlisting}[language=Python]
SJ.CalchepPartonTextInput('../data/gg_ggg_events.dat')
\end{lstlisting}

\subsection{HepMCInput}
{\bf Sample:} \verb+data/Zprime_ttbar.hepmc+

This form of input reads HepMC (version 2) format ASCII files.  This class reads in the four-vectors and PDG IDs of the particles denoted with a status code of 1 (not decayed, final state).

\subsection{PythiaInput}

This form of input generates and reads events directly from \prog{Pythia}, without ever having to write them to a file.  This requires \prog{ROOT}'s \prog{Pythia} interface; versions 6 and 8 will both work.  See \verb+examples_py/pythiaExample.py+ for an example of using \prog{Pythia} in this way.

\subsection{FourVecInput}

A \code{FourVecInput} takes four-vectors from some other code; the input class is templated on a ``Reader'' class that must provide a simple interface for retrieving four-vectors.  See \verb+IO/FourVecInput.hh+ for a complete specification, and \verb+examples_C/FourVecExample.cc+ for an example.  This does not currently work in Python, since you would have to generate a dictionary for \code{FourVecInput<YourReaderClass>}.

\subsection{Input Options}
\label{f0:inputoption} 

\subsubsection*{Multiple input files}

The \code{MultiInput} class can be used to string a set of input files together.  See \verb+examples_py/mergedInputExample.py+ for an example. For \prog{ROOT} files, the current implementation opens all input files before beginning, which may be inefficient.  Other files are opened sequentially.

\subsubsection*{Rejecting bad input}
The \code{InputMaker} can be set to remove four-vectors with negative energy and non-physical momenta by using the following function. 
\begin{lstlisting}[language=Python]
input.reject_bad_input(False)
\end{lstlisting}
The current default is \verb+False+; no checks will be done. (An alternative to this method of dealing with bad input is to use the \code{JetNegEnergyTool}, described in Sec.~\ref{f0:JetTools}.)

\subsubsection*{Reading of PDG ID codes}

The \code{InputMaker} can be set to read PDG ID codes from the input data with the following function.  This is done by default.
\begin{lstlisting}[language=Python]
input.readPdgId(True)
\end{lstlisting}
This makes the PDG IDs available for input selection and saves the IDs of the input particles for offline analysis.


\section{Output}
\label{f0:Output}

After a \SJ run, the result is a \prog{ROOT} \code{TTree} containing jet variables for each algorithm added, plus variables for input particles to the jet algorithms.  Jet and event moments are also stored.

The output \prog{ROOT} file contains a \code{TTree} (the file and \code{TTree} name are set via \code{builder.configure\_output(treename, filename)}.  The \code{TTree} contains a branch for each jet variable, named by default \{\code{AnalysisName\_eta}, \code{AnalysisName\_phi}, \code{AnalysisName\_e}, \code{AnalysisName\_mass}, \code{AnalysisName\_pt}\}.  (For this reason it is important that each analysis have a unique name!)  If input particles are being stored, there are similar branches for the input particles.  Every jet moment has an associated name and is stored in a branch named \code{AnalysisName\_momentName}.  For each analysis, an integer variable \code{AnalysisName\_N} is stored, giving the number of jets for each event.

\subsection{Output variable type}
It is possible to choose the type of the variable saved in the \code{TTree}.
The choices are C array vs.~STL vector, and floats vs.~doubles.
\begin{lstlisting}[language=Python]
# can also do SJ.kArray and SJ.kDouble
builder.set_output_type(SJ.kVector, SJ.kFloat) # (default)
\end{lstlisting}
This affects the jet momentum variables, but not jet or event moments.  Jet moments are always stored as \code{vector<T>} where \code{T} was the template parameter to the jet moment (the type that the moment returned.  Event moments are simply stored as bare \codes{T}.

\subsection{Constituents}
For all analyses, constituent information is saved as follows, assuming that the analysis was added to the \code{JetBuilder} with option \code{withIndex=True} (this is the default).  Assuming an analysis named \code{MyJet} has been added, two additional variables are stored in the \prog{ROOT} \code{TTree}:

\code{MyJet\_numC}

\code{MyJet\_ind}

\code{MyJet\_numC} is an array of size \code{MyJet\_N}.  \code{MyJet\_numC[i]} is the number of constituents of \code{i}th jet.  \code{MyJet\_ind} is an array of size \code{InputJet\_N}.  \code{MyJet\_ind[i]} is the index of the jet to which the \code{i}th input constituent has been assigned.

For example:
\begin{lstlisting}
myTree.Draw('InputJet_e', 'AntiKt10_ind==0')
\end{lstlisting}
will give the energy distribution of constituents in jet number 0 (i.e. highest $p_T$ jets) for the \code{AntiKt10} collection.

\section{Graphical interface}
\label{f0:GUI}

\begin{figure}[htbp]
\begin{center}
\includegraphics[width=1.2\linewidth]{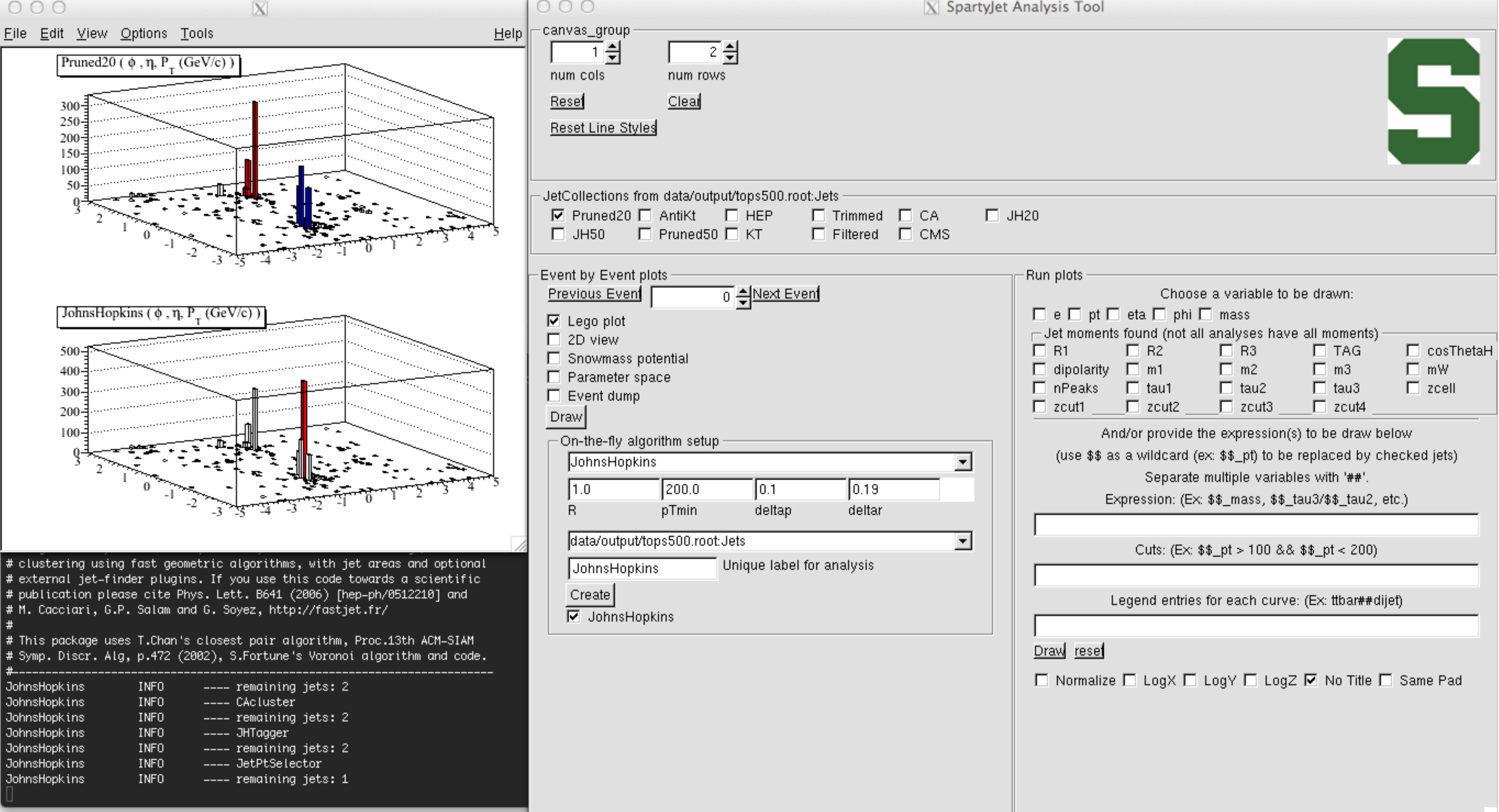}
\end{center}
\caption{A screenshot of the \SJ graphical interface.}
\label{fig:GUI}
\end{figure}

\SJ output \prog{ROOT} files can be explored with a graphical interface, launchable by running
\begin{lstlisting}
$> sparty foo.root [bar.root ...]
\end{lstlisting}
(\verb+spartyjet/bin+ must be in your \verb+$PATH+.)  A screenshot of the GUI in action is shown in Fig.~\ref{fig:GUI}.  The user can select which algorithms to view by ticking the boxes under ``JetCollections''.  On the left are several event-by-event views, which will be drawn separately for each algorithm.  The number of rows and columns are set in the upper left.

Additional analyses can be run (event-by-event) on the fly using the menu in the lower left.  These are based on a new ``live algorithm'' facility in the \SJ Python package, which extends certain common analyses with knowledge of their input parameters so the GUI can build them live with user-provided parameters.  These analyses can be run on the inputs from any file loaded on launch.  In particular, you can launch the GUI with one or more \prog{ROOT} files with no analyses run at all, only inputs.  On-the-fly analyses provide a simple and powerful way to explore jet analysis dynamics, and we hope to extend these capabilities in future releases.  High on our wish list is event generator integration, which would make the GUI self-contained.

\begin{figure}[htbp]
\begin{center}
\includegraphics[width=.7\textwidth]{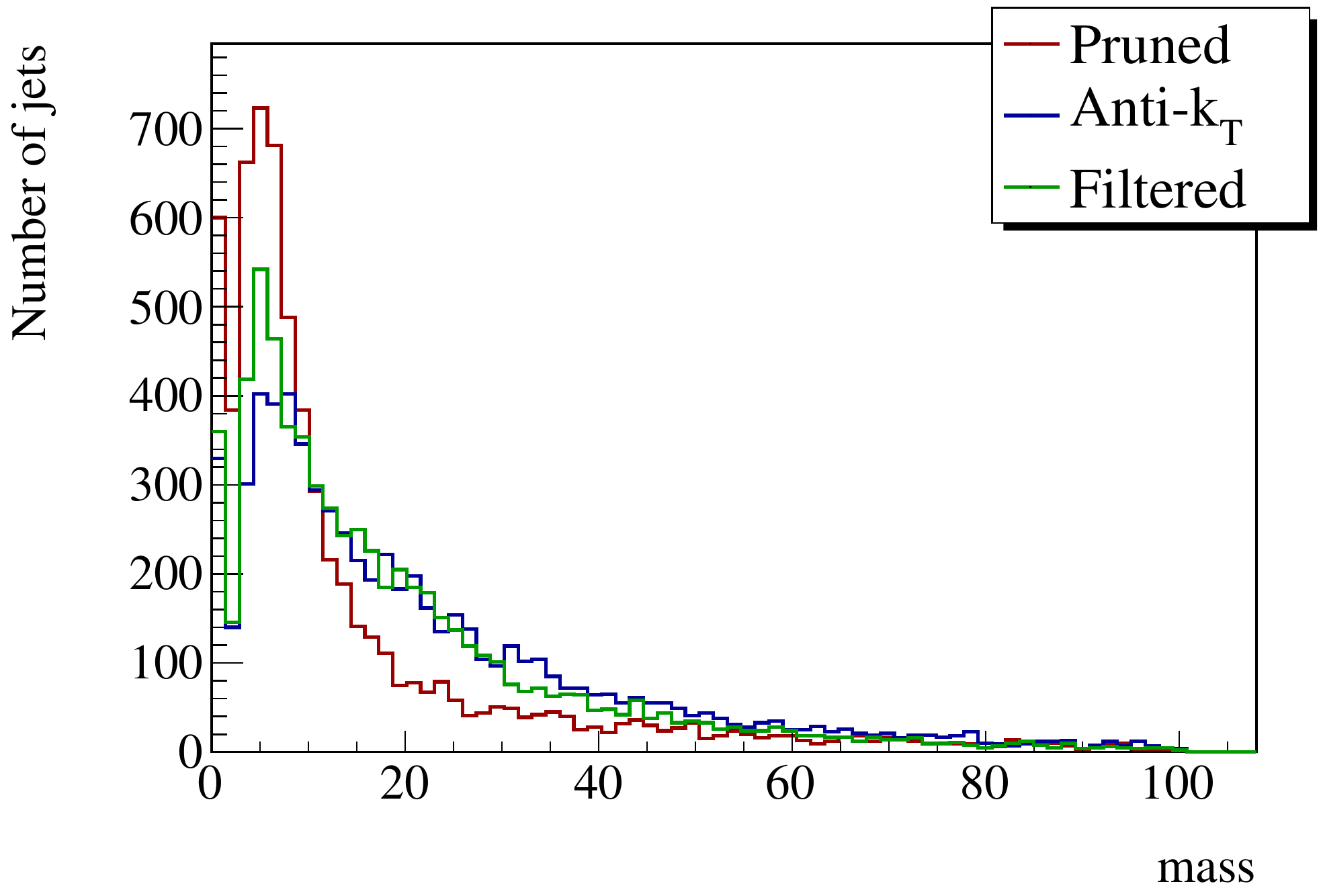}

\includegraphics[width=1.3\textwidth]{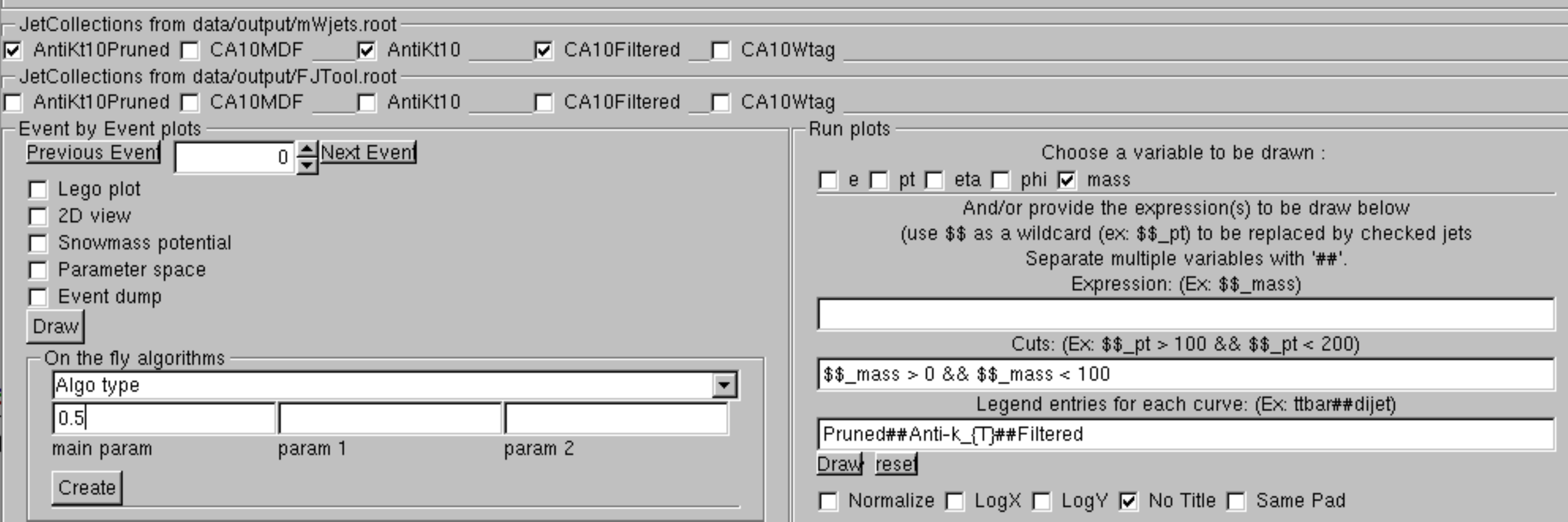}
\end{center}
\caption{An example canvas produced by the run plot option (above), with the options that produced it (below).}
\label{fig:RunPlot}
\end{figure}

On the right side of the control panel, full-run plots can be selected, which are plotted for all algorithms together (an example of this output is shown in Fig.~\ref{fig:RunPlot}).  Check boxes are provided for the standard four-momentum variables, but any stored jet moment can also be plotted by entering the name in the box below, e.g. \verb+$$_tau3/$$_tau2+ to plot the ratio of 3- to 2-subjettiness if the \code{Nsubjettiness} moment has been measured as in \verb+Boost2011.py+.  Note that \verb+$$+ is a placeholder for the analysis name, and the syntax is that of \code{TTree::Draw}.  Moments found in the input files are also given in the box below the jet variables, but note that not all moments are stored for all analyses!

Cuts can be added using \code{TCut} syntax, e.g. \verb+AntiKt10_mass > 150 && AntiK10_mass < 200+.  Finally, legend labels for each algorithm can be given; the syntax is \prog{ROOT} \code{TLatex} text, e.g. \texttt{\#phi\_\{0\}} produces $\phi_0$.  Several drawing options are provided below the \code{Draw} button.  Finer control can be gained by opening the \prog{ROOT} histogram editor in the output canvas.

\pagebreak

\section*{}
\subsection*{Contact information}

\SJ website: \url{http://projects.hepforge.org/spartyjet}\\

\underline{\bf Any questions/comments/suggestions, email:}

\begin{tabular}{l l}
Pierre-Antoine Delsart:& delsart@\,in2p3.fr\\
Joey Huston:& huston@\,pa.msu.edu\\
Brian Martin:& marti347@\,msu.edu\\
Chris Vermilion:& verm@\,uw.edu\\
\end{tabular}

\bibliographystyle{utphys}
\bibliography{sparty}

\end{document}